\PassOptionsToPackage{unicode=true}{hyperref} 
\PassOptionsToPackage{hyphens}{url}
\PassOptionsToPackage{dvipsnames,svgnames*,x11names*}{xcolor}
\documentclass[12pt]{scrartcl}
\usepackage{lmodern}
\usepackage{amssymb,amsmath}
\usepackage[T1]{fontenc}
\usepackage[utf8]{inputenc}
\usepackage{textcomp} 
\usepackage{booktabs}%
\usepackage{pifont} 
\IfFileExists{upquote.sty}{\usepackage{upquote}}{}
\IfFileExists{microtype.sty}{%
\usepackage[]{microtype}
\UseMicrotypeSet[protrusion]{basicmath} 
}{}
\usepackage[right]{lineno}
\usepackage{xcolor}
\usepackage{hyperref}
\hypersetup{
            pdftitle={Model Predictive Control on the Neural Manifold},
            pdfauthor={Christof Fehrman and C Daniel Meliza},
            colorlinks=true,
            linkcolor=blue,
            filecolor=Maroon,
            citecolor=Blue,
            urlcolor=Blue,
            breaklinks=true}
\usepackage[letterpaper, margin=1.375in]{geometry}
\ifx\paragraph\undefined\else
\let\oldparagraph\paragraph
\renewcommand{\paragraph}[1]{\oldparagraph{#1}\mbox{}}
\fi
\ifx\subparagraph\undefined\else
\let\oldsubparagraph\subparagraph
\renewcommand{\subparagraph}[1]{\oldsubparagraph{#1}\mbox{}}
\fi
\usepackage[margin=10pt,labelfont=bf,labelsep=period]{caption}
\setcapindent{0em}
\setkomafont{captionlabel}{\bfseries}

\makeatletter
\renewcommand*{\@maketitle}{%
  \global\@topnum=\z@
  \setparsizes{\z@}{\z@}{\z@\@plus 1fil}\par@updaterelative
  \begin{flushleft}%
    {\usekomafont{title}{\Large \@title \par}}%
    \vskip .5em
    {\ifx\@subtitle\@empty\else\usekomafont{subtitle}\@subtitle\par\fi}%
    \vskip .5em%
  \end{flushleft}%
    \vskip 1em%
  \par
}%
\usepackage{mathpazo}
\usepackage[doublespacing]{setspace}
\usepackage[numbers,sort&compress]{natbib}
\usepackage{graphicx}
\usepackage{xspace}
\DeclareMathOperator*{\argmin}{arg\,min}
\title{Model Predictive Control on the Neural Manifold}

\begin{document}
\maketitle

\paragraph{Authors:} Christof Fehrman\textsuperscript{1} and C Daniel Meliza\textsuperscript{2,3}

\paragraph{Affiliations:} \textsuperscript{1} Department of Mechanical Engineering and Materials Science, Duke University, Durham NC 27708, USA \textsuperscript{2}Department of Psychology, \textsuperscript{3} Program in Fundamental Neuroscience, University of Virginia, Charlottesville VA 22904, USA

\noindent
\paragraph{\textsuperscript{*}Corresponding author:} christof.fehrman@duke.edu


\bigskip
\nolinenumbers 
\section*{Abstract}
Neural manifolds are an attractive theoretical framework for characterizing the complex behaviors of neural populations. However, many of the tools for identifying these low-dimensional subspaces are correlational and provide limited insight into the underlying dynamics. The ability to precisely control the latent activity of a circuit would allow researchers to investigate the structure and function of neural manifolds. We simulate controlling the latent dynamics of a neural population using closed-loop, dynamically generated sensory inputs. Using a spiking neural network (SNN) as a model of a neural circuit, we find low-dimensional representations of both the network activity (the neural manifold) and a set of salient visual stimuli. The fields of classical and optimal control offer a range of methods to choose from for controlling dynamics on the neural manifold, which differ in performance, computational cost, and ease of implementation. Here, we focus on two commonly used control methods: proportional-integral-derivative (PID) control and model predictive control (MPC). PID is a computationally lightweight controller that is simple to implement. In contrast, MPC is a model-based, anticipatory controller with a much higher computational cost and engineering overhead. We evaluate both methods on trajectory-following tasks in latent space, under partial observability and in the presence of unknown noise. While both controllers in some cases were able to successfully control the latent dynamics on the neural manifold, MPC consistently produced more accurate control and required less hyperparameter tuning. These results demonstrate how MPC can be applied on the neural manifold using data-driven dynamics models, and provide a framework to experimentally test for causal relationships between manifold dynamics and external stimuli.

\section{Introduction}\label{Introduction}
Neural circuits are composed of large numbers of interconnected neurons whose activity depends on other neurons in the circuit. Because of these dependencies, simultaneously recorded populations of neurons usually exhibit high levels of correlation. Equivalently, most of the variance within the high-dimensional space corresponding to the firing rates of individual neurons is confined to a lower-dimensional subspace, or neural manifold \citep{cunningham2014dimensionality}. Due to their relative simplicity, neural manifolds have become a popular framework for understanding the complex dynamics of large neural populations. In many different systems, activity on neural manifolds has been shown to correlate with salient features of stimuli, physical position, and internal cognitive states \citep{mante2013context,kim2017ringdynamics,chaudhuri2019intrinsic,chung2021neural}. However, the insight that manifolds can provide into the underlying dynamics and computational principles of the circuits remains a contentious question.

Broadly speaking, neural manifolds may be seen either as descriptive tools for dealing with the inherently correlated nature of neural data arising from highly interconnected circuits, or as a method of revealing a more fundamental dynamics that exists within a latent space \citep{langdon2023unifying}. Using linear subspaces with autonomous dynamics as an illustrative example, the descriptive perspective can be seen as classic dimensionality reduction with
\begin{equation}
    \mathbf{z_t} = \mathbf{G x_t}, \label{eq:pca}
\end{equation}
where $\mathbf{x_t}$ is a column vector of the activity of $n$ neurons at time $t$ and $\mathbf{z_t}$ is a reduced dimension representation of $\mathbf{x_t}$ given by the linear transformation $\mathbf{G}$ (which parameterizes the neural manifold). This model is descriptive because the latent trajectories $\mathbf{z_t}$ are seen as a convenient representation of high-dimensional trajectories that result from the highly coupled dynamics that exist among the individual neurons $\mathbf{x_t}$, and thus provide limited information for inference or mechanistic understanding \citep{langdon2023unifying}. In contrast, the generative perspective can be modeled as a latent factor model of the form
\begin{equation}
    \mathbf{x_t} = \mathbf{F z_t} + \boldsymbol{\epsilon_t}, \label{eq:factor}
\end{equation}
where $\mathbf{x_t}$ is a column vector of the activity of $n$ neurons at time $t$, $\mathbf{z_t}$ are the latent factors that span the neural manifold with some smaller dimension $k$, $\mathbf{F}$ are the factor loadings, and $\boldsymbol{\epsilon_t}$  is a sample from some distribution (often Gaussian).  This perspective views the measured neural activity as  a function of the latent dynamics of $\mathbf{z_t}$, which emerge from but are simpler than the dynamics of $\mathbf{x_t}$. The descriptive and generative perspectives may also be seen as bottom-up and top-down approaches, respectively, for addressing the question of how computations are implemented by neural circuits \citep{gallego2017neural,langdon2023unifying}. Early work on neural manifolds used linear methods such as principal components analysis (as in equation \ref{eq:pca}) and factor analysis (as in equation \ref{eq:factor}), but there is now a broad consensus that neural manifolds are often nonlinearly embedded in the full state space, requiring more sophisticated methods to identify \citep{fortunato2023nonlinear}.

A potential weakness in the generative approach to understanding neural manifolds is that most methods of dimensional reduction are static: they produce a time series of snapshots from an informative angle in the neural state space \citep{pang2016dimensionality}, but the dynamics have to be inferred through other means. In contrast to the bottom-up approach where there is a robust foundation of biophysics on which to build models of circuit dynamics at the level of cells and synapses, the question of how best to model dynamics in the latent space remains an active area of research \citep{florian2011hidden,lusch2018deep,susillo2015neural_solution_muscle,linderman2017slds}. Testing these models and their causal relationship to behavior would benefit from methods for experimentally controlling the activity on the neural manifold. 

In this study, we develop a framework for controlling latent dynamics in the context of a sensory system. We express activity on the neural manifold with a general state-space model
\begin{align}
    \mathbf{z_{t+1}} & = g(\mathbf{z_t},\mathbf{u_t},\boldsymbol{\epsilon_t}) \label{eq:latent_dynamics} \\
    \mathbf{x_t} &= f(\mathbf{z_t}), \label{eq:obs_function}
\end{align}
where the latent dynamics on the manifold are defined by a function of the current state of the system $\mathbf{z_t}$, an external, time-varying stimulus $\mathbf{u_t}$ that will be used for control, and an intrinsic, uncontrolled source of noise $\boldsymbol{\epsilon_t}$. The high dimensional measured neural activity $\mathbf{x_t}$ is obtained with the observation function in equation (\ref{eq:obs_function}). Interestingly, we show that the data-generating process (i.e., the dynamics of $\mathbf{x_t}$) does not need to be the same as the latent dynamics model (LDM) for the framework to be useful (see Methods) and that control of a highly nonlinear neural system is possible even when agnostic to the true structure of the neural manifold.

\subsection{Controlling System Dynamics}

Given the latent dynamics specified in equation (\ref{eq:latent_dynamics}), the control problem is to find an external stimulus $\mathbf{u}$ such that the time evolution follows a specified trajectory $\mathbf{z^*}$. The field of control theory provides a rich mathematical background to find system inputs to achieve desired system outputs. Many techniques exist with feedback (closed-loop) methods being particularly attractive due to their ability to correct for unknown perturbations to the system. Broadly speaking, feedback controllers can be categorized as being either reactive or anticipatory. Reactive controllers can use present and past errors in state tracking (the difference between $\mathbf{z^*}$ and $\mathbf{z}$) to find a control signal $\mathbf{u}$. Proportional-integral-derivative (PID) control is a popular implementation of a reactive controller due to its computational simplicity and strong performance. The control input is given by 
\begin{equation}
    \mathbf{u}(t) = \mathbf{K_p e}(t) + \mathbf{K_I}\int_0^t\mathbf{e}(\tau)d\tau + \mathbf{K_D}\frac{d\mathbf{e(t)}}{dt},
    \label{eq:pid}
\end{equation}
where $\mathbf{e}(t)$ is the state error at time $t$ with gain matrices $\mathbf{K_P}$, $\mathbf{K_I}$, and $\mathbf{K_D}$. A discrete-time version of equation (\ref{eq:pid}) is easily obtained by replacing the integral term with a sum and derivative term with finite differences. Although good performance can typically be guaranteed only for linear systems, PID is still a standard method for controlling nonlinear neural systems, such as in whole-cell voltage clamp experiments.

A notable issue with reactive controllers is that corrections can only be made for errors in state tracking once they have occurred. This can be unacceptable if certain errors in state tracking are associated with pathological states. For example, suppose that a particular region of neural state space corresponded to epileptic firing. A reactive controller would only be able to respond after the system entered this region, at which point it might be more difficult to re-establish control. In contrast, anticipatory controllers are designed to predict future errors, which can allow them to prevent the system from ever entering undesirable regions of state space. Model predictive control (MPC) is an anticipatory controller that uses a model of the system dynamics to make predictions on how present and future inputs will affect errors in state tracking \citep{rakovic_handbook_2019}. Additionally, MPC is a type of optimal controller because it attempts to find a control input sequence $\mathbf{u_{1:T}}$ that minimizes a loss function of the form
\begin{equation}
    J(\mathbf{x_0}) = \sum_{i=0}^{T} \ell (\mathbf{x_i},\mathbf{u_i}), \\
    \label{eq:MPC_loss}
\end{equation}
\noindent
with constraints
\begin{align}
    \mathbf{x_{n+1}} & = f(\mathbf{x_n},\mathbf{u_n})\nonumber \\
    \mathbf{x_{LB}} & \le \mathbf{x} \le \mathbf{x_{UB}}\nonumber \\
    \mathbf{u_{LB}} & \le \mathbf{u} \le \mathbf{u_{UB}}\nonumber,
\end{align}
where $\mathbf{x_0}$ is the value of the state at the current time step and  $\ell (\mathbf{x_i},\mathbf{u_i})$ is the loss associated with $i$th time step (relative to $\mathbf{x_0}$), which is a function of either the state variable(s) $\mathbf{x}$ and input(s) $\mathbf{u}$ or both. The controller uses the dynamical model to simulate $T$ time steps into the future. Many types of loss functions are possible, but typically involve the state error and energy cost of the control signal. The constraints allow one to specify the dynamics of the system and to give lower and upper bounds for the state variables and inputs. More sophisticated versions of MPC allow for additional constraints where knowledge of any measurement or process noise can be incorporated \citep{hewing_learning-based_2020}. Although MPC is only guaranteed to be globally optimal for linear systems with convex loss functions, it can also be used in many nonlinear systems \citep{rakovic_handbook_2019,brunton_data-driven_2019}, in part because the controller can correct for model errors. 

Only the first value of the sequence $\mathbf{u_{1:T}}$ is used as input into the system, with the optimization performed again at the next time step. By repeatedly solving this optimization problem and only using the first value, the controller can correct for errors in system modeling and anticipate future changes in the desired state trajectory. This anticipation of dynamics can result in better controller performance compared to PID and other reactive controllers \citep{brunton_data-driven_2019}.

MPC has two major practical issues that are relevant to control of neural dynamics. The first is that the optimization procedure is computationally expensive and can result in poor controller performance if the time steps between measurements are small \citep{rakovic_handbook_2019}. Neural recordings are often of high dimension (e.g., extracellular recordings with high-density silicon probes) and evolve at fast time scales. The stimuli (corresponding to the input $\mathbf{u}$) may also be of a high dimension, which will result in even more complexity in optimizing equation (\ref{eq:MPC_loss}). The utility of modeling activity on the neural manifold as a latent generative process is that we can reduce the computational complexity by solving the optimization problem in a lower-dimensional state space $\mathbf{z}$. If we used the descriptive approach as in equation (\ref{eq:pca}), we would need to optimize in the original measurement dimension $\mathbf{x}$ to force the activity on the manifold to follow a specified reference trajectory. 

The second issue with applying MPC to neural systems that it requires a dynamical model of the system \citep{schwenzer_review_2021}. Although there are many dynamical models rooted in biology for individual neurons, the putative latent dynamics of a neural manifold are an emergent property that is difficult to model from first principles. This requires us to take a data-driven approach, where unknown parts of the system can be modeled via function approximation and used to predict the time-evolution of the system in response to various inputs. 

Fitting these models is achieved by observing a temporal sequence of the state and input variables with some sampling period $\Delta t$,
\begin{equation}
    \mathbf{Z} = [\mathbf{z_0},\mathbf{z_1},...,\mathbf{z_N}], \mathbf{U} = [\mathbf{u_0},\mathbf{u_1},...,\mathbf{u_N}],
\end{equation}
where $\mathbf{z_n} = \mathbf{z}(n\Delta t)$. A discrete-time model can be parameterized such that
\begin{equation}
    \mathbf{\hat{z}_{n+1}} = f_{\theta}(\mathbf{z_n},\mathbf{u_n})
\end{equation}
These type of models are often referred to as forecasting models, because the model predicts how the system will change across time. The goal is to find a set of parameters $\theta$ such that given some initial state value $\mathbf{z_0}$ and some known input $\mathbf{U}$,
\begin{equation}
    [\mathbf{z_0}, \mathbf{\hat{z}_1},...,\mathbf{\hat{z}_N}] \approx [\mathbf{z_0},\mathbf{z_1},...,\mathbf{z_N}]
\end{equation} 
for any general temporal data sequence produced from the true dynamical system.

Data-driven approaches have been successfully applied to MPC problems in diverse fields \citep{bieker_deep_2019,kaiser_sparse_2018,hewing_learning-based_2020,salzmann_real-time_2023,zheng_physics-informed_2023}, and the field is rapidly growing. In order for data-driven models to be useful for MPC applications in neuroscience, these models must be able to accurately predict the states to be controlled based only on observable state measurements, be agnostic to the number of hidden states, and generalize to a control scheme where control signals may be outside the training set. These challenges are not unique to neuroscience, but are still important to consider when selecting a data-driven approach to model the system dynamics. MPC has already been successfully applied in multiple areas of neuroscience. At the individual cell level, simulated Hodgkin-Huxley neurons have been shown to be controllable via MPC using both biophysical \citep{frohlich_feedback_2005, yue_non-linear_2022} and data-driven dynamics models \citep{senthilvelmurugan_active_2023,fehrman2024nonlinear}. Additionally, both simulated and \textit{in vivo} systems of neurons have been controlled with LQR (a related optimal control method) using optogenetic stimulation as the control signal \citep{bolus_state-space_2021}. However, to the best of our knowledge, there has been no explicit attempt to apply these methods to activity on the neural manifold. 

\subsection{Study Overview}

Our goals for this study are two-fold. First, we develop a framework for explicitly controlling the latent dynamics on a neural manifold. Second, we compare the performances of PID and MPC across several control tasks. Given that PID is computationally simple and does not require an explicit model of the system dynamics, the advantages of MPC must significantly outweigh the costs in order to advocate its implementation.

We simulated neural activity with an artificial circuit (AC) using a spiking neural network (SNN) that was driven by images of handwritten digits ($\mathbf{u}$). The activity of the network was measured as if in an extracellular recording experiment where the spike times of the SNN served as neural states ($\mathbf{x}$). Modern extracellular probes are able to record from dozens to thousands of neurons simultaneously in both anesthetized and awake animals, but this is only a small subset of the neurons in a typical local circuit. Thus, in this simulation we only used a subset of the simulated neurons to fit a LDM of the whole network. We also added random synaptic noise to simulate the influence of uncontrolled spontaneous activity arising from unobserved neurons within and outside the local circuit. 

\begin{figure}
    \centering
    \includegraphics[width=\textwidth]{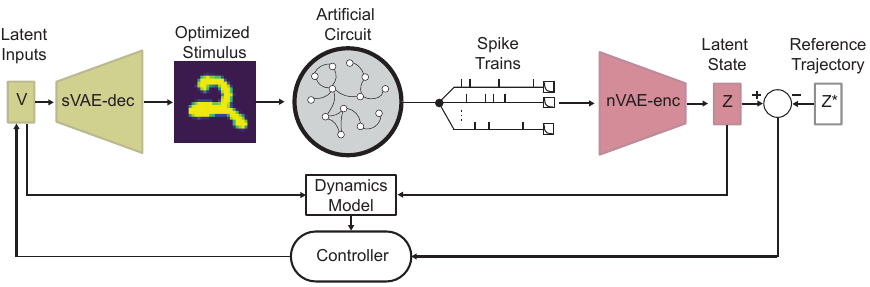}
    \caption[Control Loop of Artificial Circuit]{\textbf{Control Loop of Artificial Circuit.} Exponentially filtered spike trains are encoded into the latent state $\mathbf{z}$ and compared to a reference trajectory $\mathbf{z^*}$ to produce an error signal. MPC uses a model of the latent dynamics to find an optimal input that minimizes a loss function of the state error. This input is then projected back into the original stimulus dimension using the sVAE decoder which stimulates the artificial circuit. For PID, the path for the dynamics model is omitted and the controller uses only the present and past state errors for calculating input $\mathbf{v}$.}
    \label{fig:enter-snn-control-loop}
\end{figure}

 Using variational autoencoders (VAEs), we found low-dimensional representations of both the neural states and stimulus images (referred to as $\mathbf{z}$ and $\mathbf{v}$ respectively) and fit a linear LDM using these latent representations. Then with either PID or MPC, we found latent inputs to force the latent states on the manifold to follow specified reference trajectories $\mathbf{z^*}$. Figure \ref{fig:enter-snn-control-loop} illustrates the control loop diagram. We compared the two control strategies across two experimental conditions. First, we controlled the latent dynamics of the AC to stay at specified set points in the presence of intrinsic system noise. This result provides a proof of principle that both PID and MPC can control activity on the neural manifold. In the second experiment, we examined the generalizability of the PID and MPC controller hyperparameters from Experiment I. Our results indicated that even though MPC is more computationally complex than PID, MPC is superior across random AC subsets and proportions of measurable neurons. For a final example of neural manifold control, we used MPC to force the latent states to follow multiple time-varying reference trajectories, with the optimized visual stimuli showing striking differences. The ability to control the system to follow distinct trajectories would allow for experimenters to test for causal relationships between activity on the manifold and macro-scale behaviors.

\section{Methods}\label{Methods}
\subsection{Artificial Circuit}
\subsubsection{Architecture}
The activity of an AC evoked by an external visual stimulus was simulated with a SNN composed of three layers: sensory, reservoir, and output.
Each neuron in the AC was modeled with a recurrent leaky integrate-and-fire (rLIF) model, with the discrete time approximation
\begin{equation}
    V_{n+1} =
    \begin{cases}
        \beta V_n + w^T X_{n+1} + r^T S_n, & \text{if } V_n < \Theta \\
        0, & \text{if } V_n \geq \Theta
    \end{cases}
\end{equation}

\noindent
where
\begin{flalign*}
    &V_n: \text{membrane voltage at the \textit{n}th time step}& \\
    &\Theta: \text{spiking threshold}& \\
    &\beta: \text{decay parameter}& \\
    &X_n: \text{feedforward input vector at the \textit{n}th time step}& \\
    &w: \text{feedforward weights}& \\
    &S_n: \text{layer spiking vector at the \textit{n}th time step}& \\
    &r: \text{recurrent weights}
\end{flalign*}
Whenever the $V$ variable was reset to $0$, a spike was recorded at that time step. For a given layer with $N$ neurons, this produced a binary vector $S_n = [s_1,s_2,...,s_N]^T$ where the value of each element was either a $0$ or $1$ indicating if the corresponding neuron had fired at that time step. This allowed the activity of each neuron in a layer to be affected not only by its own firing (i.e., spiking inhibition/facilitation), but also to receive inputs from the other neurons in that layer.

Each neuron in the sensory layer received feedforward input in the form of a grayscale image reshaped into a 784-dimensional vector. This input served as the external stimulus that was the primary driver of AC activity. Gaussian noise was also added to the feedforward inputs of each neuron in every layer to produce stochasticity in activity. This noise modeled the effects of natural variability in neural firing and the effects of unknown exogenous inputs. The subthreshold activity of the three layers was given by
\begin{align}
    \text{\textbf{Sensory}}: & V_{n+1}^{sen} = \beta V_n^{sen}+w_{sen}^T(I_{n+1}+\epsilon)+r_{sen}^T S_n^{sen} \\
    \text{\textbf{Reservoir}}: & V_{n+1}^{res} = \beta V_n^{res}+w_{res}^T(S_{n+1}^{sen}+\epsilon)+r_{res}^T S_n^{res} \\
    \text{\textbf{Output}}: & V_{n+1}^{out} = \beta V_n^{out}+w_{out}^T(S_{n+1}^{res}+\epsilon)+r_{out}^T S_n^{out},
\end{align}

\noindent
where $\epsilon \sim N(0,\eta^2)$ for each element in the feedforward input and $I_n$ is the stimulus image presented at the $n$th time step. Note that $\beta$ incorporates both the membrane time constant and the time step of the discretization of the continuous LIF \citep{eshraghian2023}. The time units for the simulation can be arbitrarily chosen by modifying this parameter. For simplicity, we use units of milliseconds (ms) for each time step in all simulations, which implicitly imposes a spiking refractory period of 1 ms. Additionally, we clamped membrane voltages to a minimum value of $-2$ to prevent the membrane voltage from becoming unrealistically low. See Appendix \ref{tab:SNN_hyperparameters} for AC training and rLIF hyperparameters. 

\subsubsection{Training the Network}
In principle, the feedforward and recurrent weights for each neuron could be randomly distributed, such as in reservoir computing \citep{maass2011liquid}. However, to give the network dynamics that implement a useful computation \citep{sussillo2013opening}, the weights were trained to perform a classification task. Using the Python package \texttt{snntorch} \citep{eshraghian2023}, the AC was trained to classify digits from the MNIST data set \citep{lecun1998gradient}. Training SNNs requires additional considerations compared to traditional artificial neural networks. The resetting of a neuron's membrane voltage when it reaches the threshold $\Theta$ produces a non-differentiable function \citep{eshraghian2023} making training with gradient descent impossible. One solution to this problem is to use surrogate gradient descent \citep{neftci2019}, where the non-differentiable function is preserved in the forward pass of the network but is replaced with a sigmoid function during the backward pass. This results in a function differentiable everywhere and allows the network to be trained using standard backpropagation methods.

The MNIST data set is composed of handwritten images of the digits 0 through 9. Classification of the images was performed by associating the label of the image with a corresponding neuron in the output layer of the AC (e.g., label 4 with neuron 4). For time step $n$, the cross-entropy $\mathcal{L}_n$ was given by
\begin{equation}
    p_n^i = \frac{\exp(V_n^i)}{\sum_{k=0}^9 \exp(V_n^k)}
\end{equation}
\begin{equation}
    \mathcal{L}_n = -\sum_{i=0}^9 y_i log(p_n^i),
\end{equation}
where $V_n^i$ is the membrane voltage of the $i$th neuron which corresponds to the prediction of label $i$, and $y$ is a one-hot encoded vector of the true label. Due to the inherently temporal nature of SNNs, one must specify how many time steps a stimulus is be presented before class prediction takes place. This can be interpreted as a combination of reaction time and evidence accumulation. Thus the loss function to be minimized is given by
\begin{equation}
    \mathcal{L}_{CE} = \sum_{n=0}^T \mathcal{L}_n,
\end{equation}
where the cross-entropy loss at each time step is summed for some trial time length $T$. This forces the neuron of the associated predicted class to have the highest firing rate compared to the other neurons in the output layer \citep{eshraghian2023}. For the given time window that the image is presented, the feedforward and recurrent weights in the SNN are updated using backpropagation through time (BPTT). 

The AC was trained and assessed using half of the entire MNIST dataset (n = 35,000). This partition was split into training, validation, and testing sets using a 60/20/20 split. During training, early stopping was employed if the validation accuracy did not improve after 5 epochs. Due to the stochastic nature of the noise added to every neuron in the AC, performance after training was assessed by presenting the stimuli 30 times to obtain an accuracy distribution. Accuracy on the training (n = 21,000, M = 92.5\%, SD = 0.001\%) and validation (n = 7000, M = 90.1\%, SD = 0.001\%) set were largely similar, indicating there was little overfitting to the training data. Although the accuracies were below what would be considered competitive performance on a classification task, the purpose of training the AC was to ensure the connections between the neurons were not random.

\subsection{Dimensionality Reduction of Stimulus Set}
The high dimensionality of the stimulus inputs for the AC introduces a difficulty in the control problem. If the control input $\mathbf{u}(t)$ is found in the original measurement space, the values of $\mathbf{u}(t)$ would need to specified for each pixel value of the visual stimulus. This creates both a hyperparameter problem (the number of PID gains is equal to the number of pixels) and a computational one (MPC has to optimize over pixel space). A key insight is that the MNIST digits exhibit well-known low-dimensional structure, so we opted to find the control input in the latent stimulus space, which we denote $\mathbf{v}(t)$. Once found, the AC can be stimulated by projecting $\mathbf{v}(t)$ back into the original stimulus dimension to obtain $\mathbf{u}(t)$.

A VAE was trained to find a low-dimensional representation of MNIST digit stimuli (sVAE). The VAE framework was chosen for two reasons. First, VAEs embed high-dimensional data into nonlinear low-dimensional subspaces. This allows for a flexible approach for finding a more generalizable compression of the data compared to linear transformations \citep{gomari2022variational}. Second, the use of KL-divergence in the VAE cost function promotes nearby values of $\mathbf{v}(t)$ in the latent space to be decoded into similar values in the original stimulus space $\mathbf{U}$ \citep{kingma2013auto}. Intuitively, we want small changes in $\mathbf{v}(t)$ to produce small changes in $\mathbf{u}(t)$. If a basic autoencoder was used instead of a VAE, nearby points in latent stimulus space would not be guaranteed to be similar in the original stimulus space, which would obviously be detrimental to modeling the latent dynamics and finding suitable control inputs.

The second half of the MNIST data set (n = 35,000) was used for training the sVAE and constructing a latent input sequence for latent dynamics identification. The data set was split into training, validation, and testing sets with a 70/15/15 split. During training, if the validation loss did not improve after 5 epochs, early stopping of sVAE training was implemented. The sVAE was trained via stochastic gradient descent in \texttt{Pytorch} using the Adam optimizer with the standard VAE loss function,
\begin{equation}
    \mathcal{L}_{\text{sVAE}} =\mathcal{L}_{\text{u-recon}}+{\alpha} \mathcal{L}_{\text{v-KL}}
    \label{eq:sVAE_loss}
\end{equation}
\begin{equation}
    \mathcal{L}_{\text{u-recon}} = \sum_{i=1}^B ||\mathbf{u_i} - h_\phi^{\text{stimulus}}(\mathbf{v_i})||_2^2
\end{equation}
\begin{equation}
    \mathcal{L}_{\text{v-KL}} = \sum_{i=1}^B \sum_{j=1}^k (1+\log \sigma_{ij}^2 +-\mu_{ij}^2 - \sigma_{ij}^2), 
\end{equation}
where $B$ is the size of the batch, $h_\phi^{\text{stimulus}}$ is the sVAE decoder, $\mu$ and $\sigma^2$ are the predicted parameters of the latent Gaussian, and $k$ is the latent stimulus dimension which was chosen to be $2$. See Appendix \ref{tab:svae_architecture} and \ref{tab:sVAE_training_hyperparameters} for architecture and training hyperparameters.

\begin{figure}
    \centering
    \includegraphics{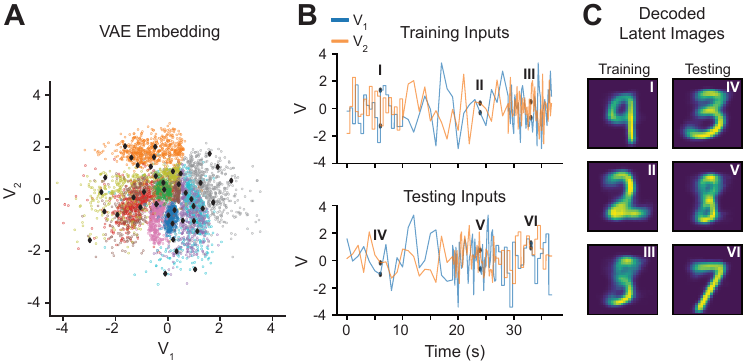}
    \caption[Latent Embeddings of MNIST Digits]{\textbf{Latent Embeddings of MNIST Digits.} \textbf{A)} Each colored dot is one of the testing MNIST digits embedded in the 2-dimensional nonlinear subspace of the sVAE encoder. Notice the clustering of the digits by label (color), indicating that digits with identical labels were often embedded in nearby latent space. In order to generate a latent sequence of inputs used to stimulate the artificial circuit, points from this latent space were sampled using \textit{k}-means clustering (black diamonds). \textbf{B)} Training and testing latent input sequences where generated using the centers of their respective latent embeddings (40 centers each) in \textbf{A}. The centers were linearly interpolated at different rates to produce a continuous latent input sequence. \textbf{C)} Three points from the training and testing inputs (black dots in \textbf{B}) decoded into the original stimulus dimension. For these time points, this is the visual input stimulus to the artificial circuit.}
    \label{fig:sVAE}
\end{figure}

A sequence of latent inputs for LDM identification was constructed using the testing set of images. Discrete points in the latent sVAE space were obtained by running \textit{k}-means clustering (n = 120) on the low-dimensional embedding of the validation images. These centers were evenly split (n = 40) into training, validation, and testing sets for LDM identification. See Figure \ref{fig:sVAE}A for latent embeddings of testing images and the corresponding centers. These discrete points were converted to a continuous time series by one of three methods: step function, fast interpolation, slow interpolation. The step function method took a sequence of centers and held each center constant for 500 time steps (ms). The fast and slow methods took the sequence and linearly interpolated values between each element in the sequence but at different time scales (200 ms for the fast and 1000 ms for the slow). The $\mathbf{V}_{\text{train}}$, $\mathbf{V}_{\text{val}}$ and $\mathbf{V}_{\text{test}}$ sequences were 36.80 seconds long (36,800 time steps). These three methods of interpolation were used to perturb the dynamics of the latent system at different timescales and frequencies. Figure \ref{fig:sVAE}B,C shows the latent input sequences and their sVAE decoded values.

\subsection{Neural Manifold and Latent Dynamics Model Identification}
The SNN was stimulated using the sVAE-decoded $\mathbf{V}_{\text{train}}$, $\mathbf{V}_{\text{val}}$ and $\mathbf{V}_{\text{test}}$ input sequences with the resulting spiking activity used to find the neural manifold and identify a LDM. For our initial experiment, the activity of a random sample of 20\% of the neurons (n = 122) in the AC was recorded. The purpose of this random sampling was to mimic the incomplete information that would be obtained in a real recording, in which only a subset of neurons are observable.

The measured binary spiking states were converted to continuous states with an exponential filter, where the smoothed state $\mathbf{x_n}$ of spiking state $\mathbf{s_n}$ is given by
\begin{equation}
    \mathbf{x_{n+1}} = \omega \mathbf{s_n} + (1-\omega)\mathbf{x_n}
\end{equation}
where $\omega$ was chosen to be $0.1$ and $\mathbf{x_0} = \mathbf{s_0}$. The neural manifold of the AC is the low-dimensional space $\mathbf{Z}$ parameterized by a latent encoding of this measured neural activity.

We can now restate our original control problem as follows: given a latent encoding of the neural states
\begin{equation}
    \mathbf{z_n} = f_{\theta}^{\text{neural}}(\mathbf{x_n}),
\end{equation}
we seek to find a latent control input $\mathbf{v_n}$ such that it forces $\mathbf{z_n}$ to follow some desired trajectory $\mathbf{z_n^*}$. This control input can be found easily for PID by scaling the present and past errors by their corresponding gains in equation (\ref{eq:pid}). For MPC, we find an optimal set of latent inputs
\begin{equation}
    \mathbf{v}_{1:T} = \argmin_{\mathbf{v}_{1:T}} \sum_{n=0}^{T} \ell(\mathbf{z_n},\mathbf{v_n})
\end{equation}
given the dynamics model
\begin{equation}
    \mathbf{z_{n+1}} = g(\mathbf{z_n},\mathbf{v_n})
\end{equation}
and any additional constraints.
For both control strategies, the AC can then be stimulated with the decoded latent inputs
\begin{equation}
    \mathbf{u_n} = h_{\phi}^{\text{stimulus}}(\mathbf{v_n})
\end{equation}
to produce the neural states $\mathbf{x_{n+1}}$ at the next time step.

There are many ways to find the latent embedding of the neural activity and the resulting latent dynamics model. One approach is to identify the neural manifold by applying standard dimensionality reduction methods to recorded activity such as PCA and autoencoders \citep{gallego2017neural} and then identify the LDM within this latent space using data-driven modeling with sequences $\mathbf{z_{0:T}}$ and $\mathbf{v_{0:T}}$. However, learning the neural manifold separately from the dynamics can lead to suboptimal performance, because the latent space may not be well-aligned with the system's underlying dynamics \citep{lusch2018deep}. Several methods have been proposed to simultaneously learn latent dynamics and structure \citep{sani2021modeling, gosztolai2025marble, schneider2023learnable,abbaspourazad2024dynamical}. A popular method is Latent Factor Analysis via Dynamical Systems (LFADS) and its related variants \citep{pandarinath2018inferring,keshtkaran2022large}, but these are based on recurrent neural networks (RNNs), which can be challenging to train and interpret. Instead, for simplicity we use a VAE-based reduction of the neural states and implement a dynamics model structure similar to LFADS but without hidden recurrent components. Our purpose here is not to innovate LDM architecture design, but to demonstrate how dimensionality reduction and data-driven dynamical systems modeling can be used to control the neural manifold. We suspect that the best model for neural manifold control will depend on the properties of the system, control apparatus used, and desired behavior.

Our approach consists of two stages: first, we pretrain a VAE to learn a structured latent representation of neural activity (nVAE). Then, we incorporate the pretrained nVAE and the sVAE into a larger dynamics model that learns linear dynamics within the latent space. While the parameters of the nVAE are free to vary during LDM training, the parameters of the sVAE are frozen, under the simplifying assumption that the latent structure of the stimulus is independent of the neural manifold's structure, though the reverse is not necessarily true.

The nVAE was pretrained using the smoothed spikes $\mathbf{X}$ to start the nVAE encoder and decoder at a good point for modeling the latent dynamics. The dataset $\mathbf{X}$ had training, validation, and testing partitions corresponding to the stimulus set that produced them. During pretraining, the nVAE loss was given by 
\begin{equation}
    \mathcal{L}_{\text{nVAE}} = \mathcal{L}_{\text{x-recon}}+{\alpha} \mathcal{L}_{\text{z-KL}}
\end{equation}
\begin{equation}
    \mathcal{L}_{\text{x-recon}} = \sum_{i=1}^B ||\mathbf{x_i} - h_\phi^{\text{neural}}(\mathbf{z_i})||_2^2
\end{equation}
\begin{equation}
    \mathcal{L}_{\text{z-KL}} = \sum_{i=1}^B \sum_{j=1}^k (1+\log \sigma_{ij}^2 +-\mu_{ij}^2 - \sigma_{ij}^2), 
\end{equation}
which had a corresponding structure to the loss function for the sVAE. The latent dimension of $\mathbf{Z}$ was chosen to be two. During pretraining, the learning rate was progressively reduced using a Lambda learning rate scheduler with a decay factor of 0.99. See Appendix \ref{tab:nVae_architecture} and \ref{tab:nVAE_pretraining_hyperparameters} for nVAE architecture and pretraining hyperparamters. 

After pretraining the nVAE for 20 epochs with $\mathbf{X_{\text{train}}}$, the full LDM was given by equations
\begin{equation}
    \mathbf{\hat{z}_{n+1}} = \mathbf{A}f_{\theta}^{\text{neural}}(\mathbf{x_n})+\mathbf{B}f_{\theta}^{\text{stimulus}}(\mathbf{u_n})
\end{equation}
\begin{equation}
    \mathbf{\hat{x}_{n+1}} = h_{\phi}^{\text{neural}}(\mathbf{\hat{z}_{n+1}})
\end{equation}.

The fine-tuning of the nVAE encoder and decoder as well as estimating unknown matrices $\mathbf{A}$ and $\mathbf{B}$ were found by minimizing the loss function 
\begin{equation}
    \mathcal{L}_{\text{LDM}} = \mathcal{L}_{\text{nVAE}}+\mathcal{L}_{\text{z-forecast}}
\end{equation}
\begin{equation}
    \mathcal{L}_{\text{z-forecast}} = \frac{1}{T}\sum_{i=0}^{T} ||\mathbf{z_{n+i+1}}-\mathbf{\hat{z}_{n+i+1}}||_2^2
\end{equation}
where the forecasting loss term ensures that the learned latent representations evolve consistently with the observed neural dynamics. Latent forecasts were obtained by encoding the observed spike behavior at time step $n$ as the initial value and using the latent input sequence $\mathbf{v_{n:n+T-1}}$ to auto-regressively predict $T$ timesteps into the future. The model was trained until the loss from validation data $\mathbf{X_{\text{val}}}$ and $\mathbf{V_{\text{val}}}$ did not improve after 5 epochs. The learning rate was also reduced with a Lambda learning rate scheduler with a decay factor of 0.99. See Appendix \ref{tab:LDM_training_hyperparameters} for LDM fine-tuning hyperparameters.

\begin{figure}
    \centering
    \includegraphics{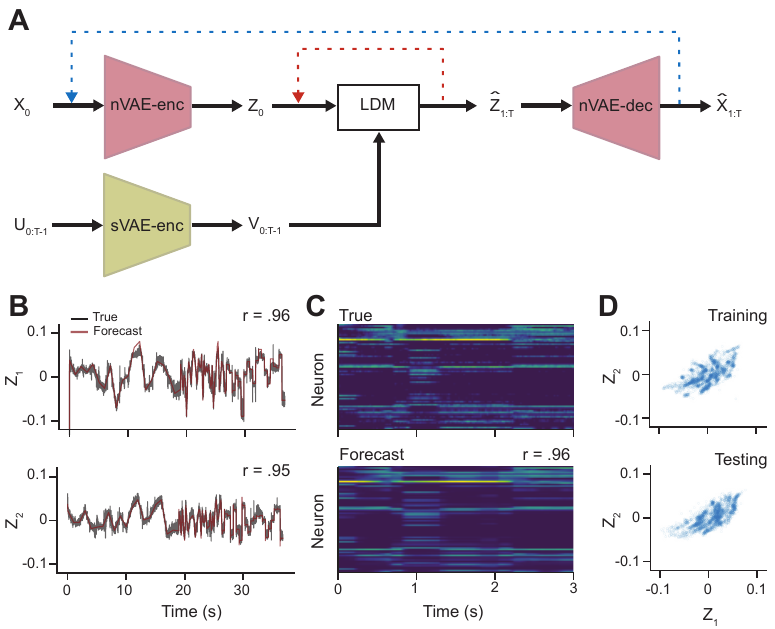}
    \caption[Forecasts and Latent Embeddings of Neural Activity]{\textbf{Forecasts and Latent Embeddings of Neural Activitiy.} \textbf{A)} Open-loop forecasting diagram for assessing the quality of LDM. For a given initial spiking state $\mathbf{x_0}$, a sequence of latent states $\mathbf{Z}$ or spiking states $\mathbf{X}$ can be predicted autoregressively for a given sequence of inputs $\mathbf{U}$. Two types of forecasting strategies are shown, all latent state prediction (red) and measurement spiking state prediction (blue). Since MPC is occurring in the latent space, predictive accuracy in the red loop is the primary measure of model performance. While we did not explicitly use the forecasts in the spike measurement space, the high correlation between the predicted and actual spikes in the test set indicates that the model captures both the measured and latent dynamics of the neural system. \textbf{B)} Open-loop forecasts (red) of the latent states $\mathbf{Z_1}$ and $\mathbf{Z_2}$ on testing data (black). \textbf{C})  A subset of the smoothed spikes from the testing data (above) and the open-loop forecasted values (below). The correlation shown is the for the entirety of the testing data set, not just for the subset shown here. \textbf{D}) Latent embeddings of the $\mathbf{Z}$ state variables obtained by applying the nVAE-enc to both the training and testing smoothed spikes. The embeddings shown here are downsampled by a factor of 10 for visual clarity.}
    \label{fig:LDM_forecasts}
\end{figure}

LDM performance was assessed by using an initial value of $\mathbf{z_0}$ and forecasting for the entire training sequence length (n = 36,799). At every time step, the known value of $\mathbf{V_{\text{test}}}$ and the model's previous prediction of $\mathbf{z}$ was used to forecast the next value (i.e., open-loop forecasting). There was a high correlation between the open-loop forecast and actual latent trajectory $Z$ ($R_{z_1} = 0.96$, $R_{z_2} = 0.95$) indicating that the dynamics model would be useful for control with MPC. See Figure \ref{fig:LDM_forecasts} for LDM forecasting performance. While PID does not require an explicit model, we compare both PID and MPC within the same learned latent space parameterized by the LDM to evaluate their relative performance.

\section{Results}\label{Results}
\subsection{Experiment I: Both PID and MPC Can Force the Latent Dynamics on the Neural Manifold to Stay at Set Points}
As a simple first experiment, the latent dynamics on the AC neural manifold were controlled to follow a step function. This reference trajectory was composed of two set points in latent space, each held constant for 500 ms. The values were chosen by running \textit{k}-means clustering on the latent state testing data and using the resulting centroids as the set points. 

The MPC controller had a predictive time horizon $T$ of 10 time steps and optimized the loss function
\begin{equation}
    J(\mathbf{z_0}) = \mathbf{z}^\intercal_\mathbf{T} \mathbf{S} \mathbf{z_T} + \sum_{i = 0}^{T-1} \mathbf{z}^\intercal_\mathbf{i} \mathbf{Q} \mathbf{z_i} + \Delta \mathbf{v}^\intercal_\mathbf{i} \mathbf{R} \Delta \mathbf{v_i}
\end{equation}
where
\begin{equation}
\mathbf{Q,S} = 
\begin{bmatrix}
1 & 0 \\
0 & 1
\end{bmatrix},
\mathbf{R}= 
\begin{bmatrix}
0.001 & 0\\
0 & 0.001
\end{bmatrix}.
\end{equation}
The values in the loss function matrices were hand-tuned and required fewer than a dozen runs to produce good performance. No constraints other than the dynamics model were included. 

Because the latent control inputs $\mathbf{v_n}$ were two-dimensional, the PID gains in equation (\ref{eq:pid}) were all 2$\times$2 matrices. Due to the difficulty in hand-tuning all the PID gains, a grid search of the PID $\mathbf{K_P}$ gains was performed across each dimension with $\mathbf{K_I}$ and $\mathbf{K_D}$ set to $\mathbf{0}$. There were 6 candidate values for each element of $\mathbf{K_P}$: $[0, 0.1, 1, 10, 20, 50]$. Using the values that resulted in the lowest error in state tracking, we then manually tuned the value of $\mathbf{K_I}$ and $\mathbf{K_D}$. The values of the PID gain matrices used in all experiments were
\begin{equation}
\mathbf{K_P} = 
\begin{bmatrix}
20 & 20 \\
5 & 20
\end{bmatrix},
\mathbf{K_I} = 
\begin{bmatrix}
0.9 & 0.9 \\
0.1 & 0.1
\end{bmatrix},
\mathbf{K_D}= 
\begin{bmatrix}
0.01 & 0.01\\
0.01 & 0.01
\end{bmatrix}.
\end{equation}

Due to the stochastic nature of the AC responses, 50 independent trials of MPC and PID were performed. All MPC optimizations and implementations were performed using the \texttt{do-mpc} python package \citep{fiedler_-mpc_2023}. This package utilizes CasADi \citep{andersson_casadi_2019} and IPOPT \citep{wachter_implementation_2006} for interior-point optimization and automatic differentiation methods.

\begin{figure}
    \centering
    \includegraphics[width=\textwidth]{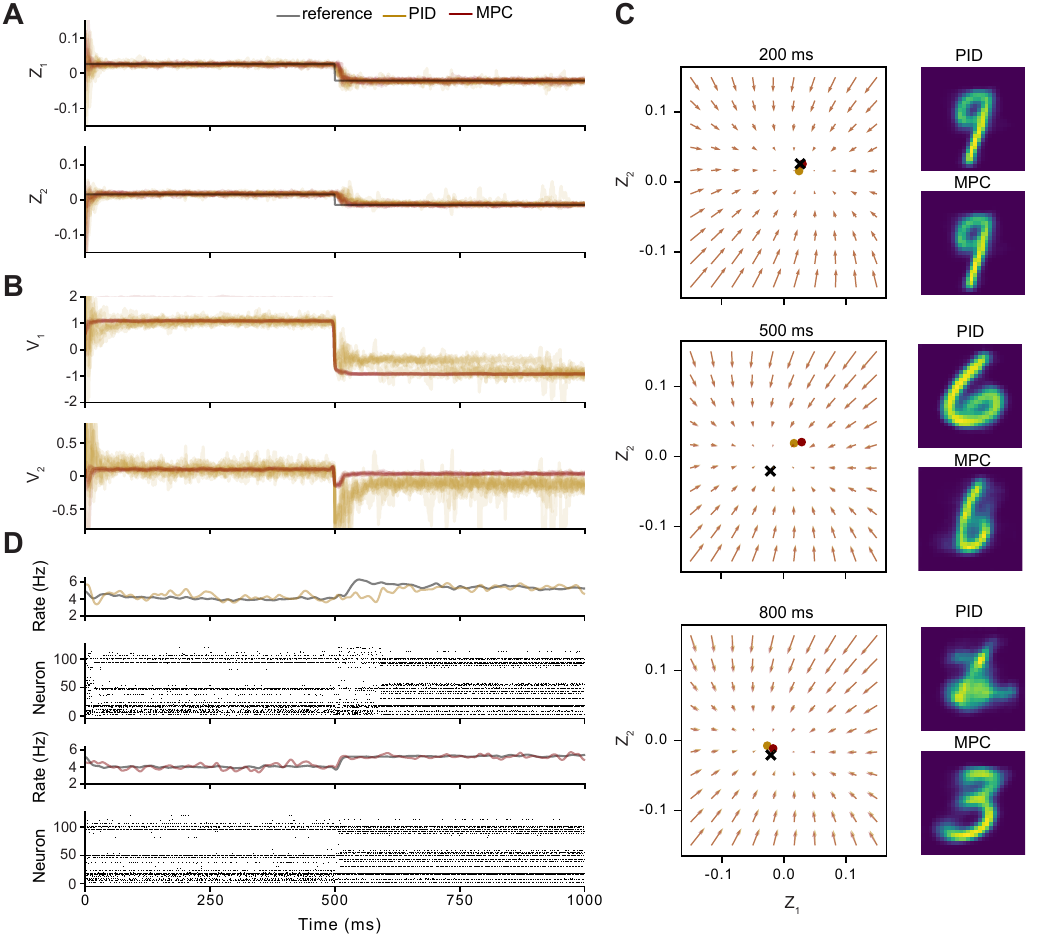}
    \caption[Set Point Control Comparison between PID and MPC]{\textbf{Set Point Control Comparison between PID and MPC}. \textbf{A)} The reference trajectory (black) was composed of two set points that changed at 500 ms. In color are the controlled latent states from both PID (yellow) and MPC (red) across fifty trials. \textbf{B)} Corresponding control inputs obtained from each control method. \textbf{C)} At each time step, the LDM predicts a vector field in $Z$ state space. Three snapshots of this vector field are shown at 200, 500, and 800 ms. The desired reference points are shown by a black X with the controlled latent states shown for both PID (yellow) and MPC (red). Images on the right are the decoded latent inputs from the optimizer at each timepoint. \textbf{D)} Rasters and firing rates of measured spikes across control methods for a single trial. For each raster plot, the population mean firing rate for one trial is shown above in color, and the average across all 50 trials is shown in black. An animated visualization of the latent state and the stimuli used for control can be found in Supplementary Material V1: \url{https://doi.org/10.6084/m9.figshare.28674500.v1}}
    \label{fig:exp_1}
\end{figure}

Both controllers were able to achieve good performance, especially when considering the noise in the system and that only 20\% of the neurons of the AC were observable. Even though the noise that was added to all neuron inputs was Gaussian, the nonlinearities in the rLIF models propagate highly complex noise structures throughout the network. At the beginning of control and after the set point was switched, the system converged to the desired location in the latent space within 20--50 time steps and then largely remained there despite continual noise inputs (Figure \ref{fig:exp_1}A). However, the PID control input had more variability between trials, suggesting that MPC was better able to account for the unknown full-state system dynamics and sources of noise. 

Because the latent space only had two dimensions, we can visualize the dynamics of the forecasting model as a vector field and the current and desired states of the system as points within this field. The dynamics can be decomposed into an autonomous component $\mathbf{A z_n}$ and the forcing from the stimulus $\mathbf{B v_n}$, and as the combined system is linear, there is at most a single equilibrium. As seen in Figure \ref{fig:exp_1}C, at time points when the set point was constant, both controllers chose stimuli such that the dynamics had a stable equilibrium close to the set point. While the control input $\mathbf{v}$ was highly consistent between trials for MPC, there was greater variability for the PID control inputs. In an experimental setting, this would result in less ability to distinguish the effects of the latent input from neural manifold dynamics.

As seen in Figure \ref{fig:exp_1}B, there was a qualitative change in the activity of the population when the reference point changed at 500 ms for both controllers. We can also examine the stimuli that were actually presented to the AC during control. As seen in Figure \ref{fig:exp_1}C, nearly all the control stimuli had digit-like characteristics, which undoubtedly reflects the fact that the sVAE was trained only on digits.

\subsection{Experiment II: MPC Hyperparameters Generalize Better to New Experimental Conditions Compared to PID}
To investigate how generalizable the controller hyperparameters were to different experimental conditions, the procedure from Experiment I was performed on different subsets of neurons randomly sampled from the same AC. If the controllers from Experiment I did not generalize to other samples, there would be considerable time cost to performing this experiment in a laboratory setting. For each level of neuron observability, 10 independent ensembles were sampled from the SNN (n = 80). A LDM was trained for each ensemble, but in order to ensure a fair comparison, the architecture was kept constant (except for the size of the input layer). The nVAE and LDM training hyperparameters were kept the same as in Experiment I. The normalized mean square error (nMSE) was used to quantify the controller performance of the latent state, which normalizes mean square error by the difference between the maximum and minimum $\mathbf{z^*}$ values. This was done for ease of comparing controller performance across dimensions with different scales.

\begin{figure}
    \centering
    \includegraphics[scale=0.8]{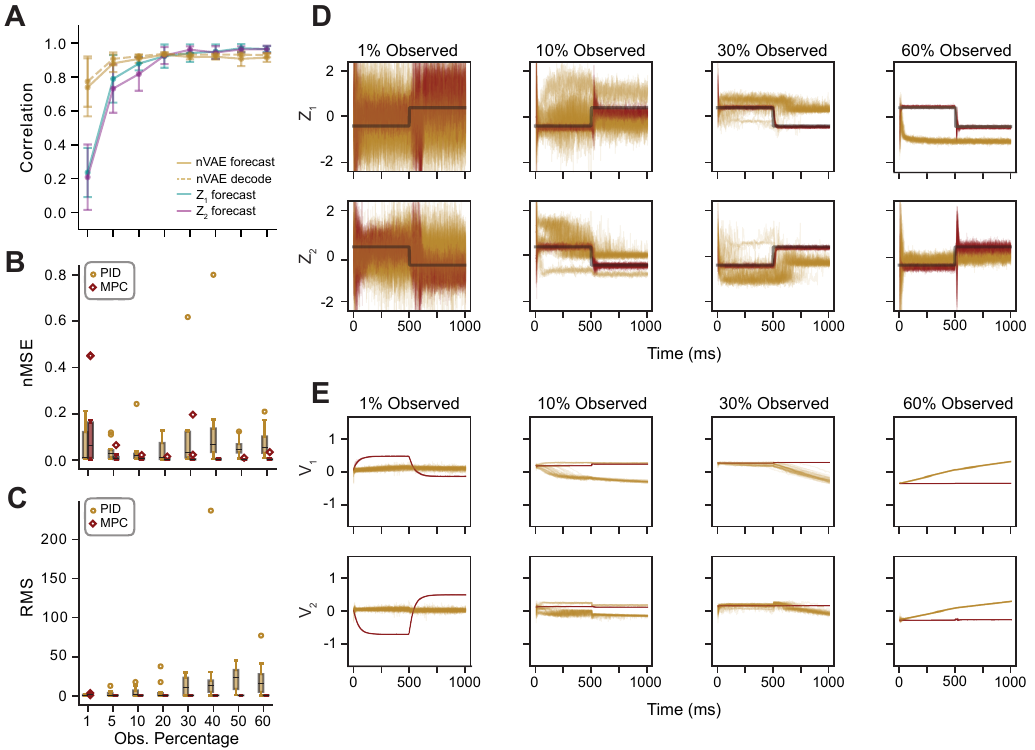}
    \caption[Transferability of control hyperparameters across experimental conditions. ]{\textbf{Transferability of control hyperparameters across experimental conditions.} \textbf{A)} The forecasting performance of LDMs is evaluated based on the correlation between the testing data and predicted open-loop forecasts, as well as the correlation between smoothed spikes and nVAE forecasted spikes or decoded $\mathbf{z}$ forecast (solid and dashed yellow lines, respectively). For each observation condition, 10 independent SNN ensembles were sampled, and the average performance across ensembles is plotted for each observation percentage. \textbf{B}. The average nMSE of the latent states across ensembles for the PID (yellow) and MPC (red) control methods. For all conditions except for the 1\% observability condition, the MPC nMSEs were smaller, demonstrating that MPC did a better job of controlling the latent states. The controller hyperparameters for both methods were tuned only for a single ensemble in the 20\% observability condition. The relatively stable nMSEs for MPC indicates that the controller hyperparameters are transferable to different experimental conditions. \textbf{C)} Same as \textbf{B}, but the average RMS of the control input $\mathbf{v}$. Notice that as the observability increases, so does the variance of the average RMS for PID inputs, but not MPC inputs. \textbf{D)} Exemplar latent trajectories for set point control for both PID (yellow) and MPC (red) control methods across fifty trials. Ranges have been scaled relative to the reference trajectories. \textbf{E}) Corresponding latent inputs for \textbf{D}. Ranges have been scaled by the maximum PID range for each condition.}
    \label{fig:exp_2}
\end{figure}

As shown in Figure \ref{fig:exp_2}A, there was a positive, monotonic relationship between the proportion of neurons observed and performance of the LDM. Both the forecasting and nVAE decoding performances were low at only 1\% observability, and there was a steep increase in performance that saturated at 20\% observability. It is likely this saturation is due to the architecture design choices being tuned at the 20\% level in Experiment I, but also indicates that there are diminishing returns to including more neurons in a recording session. To assess controller performance, the latent dynamics inferred at each level of observability were controlled to follow a step function of two set points obtained from \textit{k}-means clustering, as in Experiment I. Fifty trials were run for each of the ensembles.

Shown in Figure \ref{fig:exp_2}B, the MPC hyperparameters from Experiment I did not generalize well to the 1\% observation level, likely because of the poor LDM performance. However, as the observation percentage went up, the nMSE in MPC tracking error decreased, indicating that the hyperparameters generalized to these conditions. In contrast, the PID controller did not generalize well and had higher variability in average nMSE. As seen in \ref{fig:exp_2}B, except in the 1\% observation condition, MPC had superior performance. Not only was the latent state tracking of PID even nosier than MPC, but the PID controller could catastrophically fail, producing a controlled trajectory far from the desired set point. As seen in \ref{fig:exp_2}C, the distribution of the average root mean square (RMS) of the control input had greater variability compared to MPC. This variability in control input suggests that MPC provides a more stable mapping between inputs and latent states, making it better suited for understanding the neural manifold compared to PID. If the goal is to design experiments that associate specific inputs with specific latent states, the oscillatory and inconsistent control signals of PID (as seen in \ref{fig:exp_2}E) would make this challenging. In contrast, MPC's more stable and consistent control inputs lead to outputs that are easier to interpret in latent space. Given these results, we conclude that control over neural manifold dynamics is better achieved with MPC as opposed to PID due to its robustness to unknown noise and generalizable hyperparameters.

\subsection{Experiment III: Distinct Reference Trajectories Produce Different Levels of Controller Performance}
The previous experiments revealed how MPC used the stimulus-dependent dynamics of the forecasting model to drive the network to specific locations in the latent space. However, the forecasting model is only a linear approximation of the simulated network's dynamics, which are nonlinear and of a much higher dimension. To examine how MPC could be used to probe the underlying dynamics of the network, it is not sufficient to characterize the start and end points of a trajectory, but instead what specific path was taken. In this experiment, the reference trajectory was replaced with two time-varying functions (reference trajectory 1 and 2). Each of these reference trajectories had the same initial ($\mathbf{z_0^*}$) and final values ($\mathbf{z_f^*}$), but took different paths through latent state space. Using a parameterized function of a circle passing through points $\mathbf{z_0^*}$ and $\mathbf{z_f^*}$, trajectories 1 and 2 were the opposite arcs of the resulting circle. This ensured that both trajectories were of equal length through the latent state space. The observation model from Experiment I was used for this task (20\% observability) and the values of $\mathbf{z_0^*}$ and $\mathbf{z_f^*}$ were the set points from the same experiment. Fifty control trials were performed for each of the reference trajectories, with each trial having the same MPC hyperparameters from Experiments I and II.
\begin{figure}
    \centering
    \includegraphics{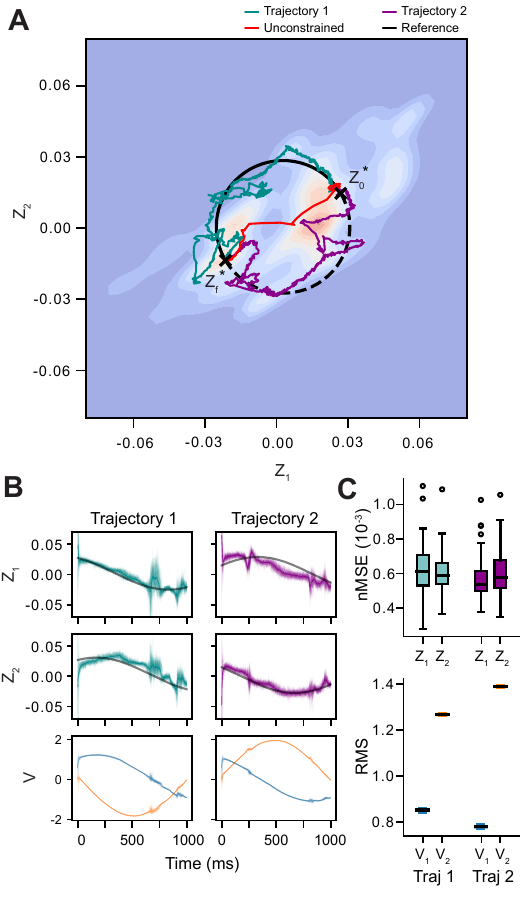}
    \caption[Latent States and Inputs for Different Reference Trajectories. ]{ \textbf{Latent States and Inputs for Different Reference Trajectories. }  \textbf{A)} Average controlled paths (cyan and magenta) for two reference trajectories (solid and dashed black lines). Transients (first 25 time steps) were removed for clarity. $\mathbf{z^*_0}$ is the initial point and $\mathbf{z^*_f}$ the final point in both trajectories. Mean MPC path from Experiment I is shown in red. Set points were identical to $\mathbf{z^*_0}$ and $\mathbf{z^*_f}$, but there was no constraint on the path taken between them. Pixel intensity shows density of observations used for fitting the LDM. \textbf{B)}   Controlled latent states (top and middle) and latent inputs (bottom) for fifty trials. Latent inputs are $\mathbf{V_1}$ (blue) and $\mathbf{V_2}$ (orange). \textbf{C)} (Above) Distribution of $\mathbf{Z_1}$ and $\mathbf{Z_2}$ nMSEs for each reference trajectory. (Below) Distribution of RMS for $\mathbf{V_1}$ (blue) and $\mathbf{V_2}$ (orange) control inputs. See Supplementary Material V2, V3 for animated visualization of the controlled trajectories: \url{https://doi.org/10.6084/m9.figshare.28674500.v1}.}
    \label{fig:exp_3_latent}
\end{figure}
\begin{figure}
    \centering
    \includegraphics[]{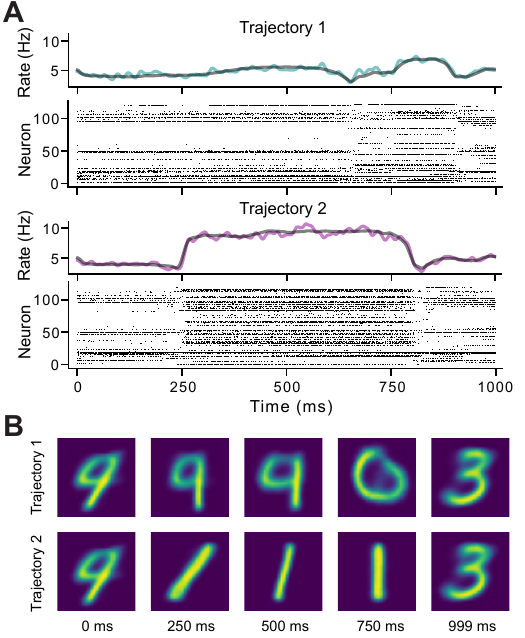}
    \caption[Recorded Spikes and Control Inputs for Experiment III]{\textbf{Recorded Spikes and Control Inputs for Experiment III}. \textbf{A)} Rasters and firing rates from a single trial for the reference trajectory 1 and 2 conditions. For each raster plot, the firing rate for the single trial is shown above in color and the average across 50 trials is shown in black. \textbf{B)} The decoded latent inputs that were used to stimulate the SNN during specific time points of the single trials in \textbf{A}.}
    \label{fig:exp_3_measurement}
\end{figure}

As seen in Figure \ref{fig:exp_3_latent}A, the controller successfully forced the network to take two different trajectories through the latent space, both of which were different from the trajectories the system used in Experiment I. There were clear differences between the two trajectories in the RMS of the latent inputs produced by the controller (Figure \ref{fig:exp_3_latent}C). In both cases, there was more ``power'' required for $\mathbf{V_2}$ than in $\mathbf{V_1}$, but the difference in RMS was much larger for trajectory 2. Examining the activity in the measurement space, there were obvious differences in the spiking behavior of the AC (Figure \ref{fig:exp_3_measurement}A). The visual stimuli produced from decoding the latent inputs were initially similar between the two reference trajectories but then diverged as the trajectories separated (Figure \ref{fig:exp_3_measurement}B). This is consistent with the behavior of the AC, which was trained to produce distinct responses to different digits; thus, different regions of the state space are likely to correspond to specific digits. One implication of this is that MPC of the latent dynamics could reveal if specific kinds of stimuli correspond to particular latent trajectories or if the activity in the latent space is driven by the differences in the stimulus at every time step (e.g. prediction error \citep{egner2010expectation}). 

\section{Discussion}\label{Discussion}
In this study, we demonstrated that both PID and MPC can be used to control the latent states of an artificial spiking neural network. MPC had superior performance, which we attribute to its anticipatory nature and explicit use of data-driven models of the latent dynamics. Despite the highly nonlinear dynamics of the AC, the latent states could be fixed to certain set points or forced to follow arbitrary, time-varying reference trajectories. Control was possible even when only a limited proportion of the neurons was observed and there when there were unknown sources of noise in the network. Reducing the dimensionality of the neural activity and the stimuli used to control the network can therefore make MPC computationally tractable for high-dimensional neural systems. Although this may seem like a convenient trick that only works in our artificial example, we hypothesize this method of finding control inputs in the latent space $\mathbf{V}$ can be extended to other experimental conditions. For example, naturalistic stimuli exhibit both linear and nonlinear correlations that can be used for dimensionality reduction in the stimulus space \citep{Goffinet:2021lowdimensional}. Based on this approach and this proof of principle, it should be possible to use MPC for real-time control of latent dynamics in sensory-driven biological networks using extracellular recordings. Achieving this level of control over neural networks would enable experimenters to probe the dynamics of neural circuits in novel ways. For example, perceptual decision boundaries have been modeled as attractor states in the neural manifold \citep{khona2022attractor}. By controlling this latent activity along specific trajectories, researchers could more precisely investigate the visual features of the control stimulus that drive decision-making. Additionally, the framework we provide here can be extended to other modalities of stimulation. For example, instead of using external visual stimuli as $\mathbf{U}$, complex spatial patterns using two-photon holographic optogenetic stimulation could be used to control latent neural activity \citep{adesnik2021probing}. This could allow for exciting new experimental designs where the relationship between organism behavior and latent neural activity could be more deeply explored. 

In Experiment II, we assessed the generalizability of MPC and PID controller hyperparameters by applying them to different ensembles of neurons randomly sampled from the same AC. Across all but the 1\% observation condition, MPC demonstrated significantly better state tracking performance than PID, despite both controllers being tuned to a single ensemble in the 20\% observation condition. Additionally, PID exhibited greater variability in its latent control inputs, which would make it more challenging to interpret the relationship between neural manifold activity and stimulus structure. One potential concern with our simulation is that recording even 1\% of the neurons in a real neural circuit would be an exceptionally high percentage. However, we hypothesize that the poor performance of MPC in the 1\% condition was primarily due to deficiencies in the learned dynamics model rather than the proportion of neurons recorded. As shown in Figure \ref{fig:exp_2}B, there were some ensembles that still resulted in very good control performance. While a detailed analysis of this issue is beyond the scope of the current study, future research should investigate how well MPC and other control systems perform in simulated circuits with different connectivity patterns and redundancy in neural responses.

In Experiment III, we found that the state of the network could be forced to take different trajectories by providing distinct sequences of latent inputs, but there were regions of the latent space where control was better or worse. For both trajectories, it appeared as though the system was attracted to specific regions of the state space, and that there were other regions the controller was unable to force the system to enter. Interestingly, individual trajectories appeared to show large oscillations in some regions of the state space, which could reflect rapid switching between stable equilibria or a limit cycle. One explanation is that these regions correspond to attractors and separatrices in the true dynamics of the AC. The potential sources of these oscillations may have been obfuscated by our use of only two latent dimensions in all of our experiments. While this was done for ease of visualization and interpretation, a more rigorous analysis to determine the dimensionality of the latent space \citep{levina2004maximum, chen2022automated} may have resulted in better forecasting and control. For example, the oscillatory behavior seen in some regions of the latent state space in Figure \ref{fig:exp_3_latent} may have been from a saddle point that would have been identified if we had used three or more latent dimensions. Another possibility is that the LDM made better predictions for certain regions of latent state space, where the dynamics of the full network were more linear. The use of a nonlinear dynamics model such as hybrid/switching linear dynamics \citep{linderman2017slds, song2022modeling}, time-varying dynamics \citep{yang2021adaptive}, or structured variational autoencoders \citep{johnson2016structured} may result in a better approximation of the latent vector field. Although this would likely improve the performance of the controller, it would also introduce greater complexity when fitting the model and in the MPC optimization. However, recent work for finding linear representations of nonlinear dynamical systems using Koopman theory offers promising evidence that a linear modeling approach can be effective for controlling neural manifold dynamics \citep{lusch2018deep, kaiser2021data}. Extensive work may be needed to identify the best approaches for balancing forecasting accuracy with computational efficiency for a given system and set of scientific questions.

An important practical consideration is how to decide if either controller is good enough for a particular application. We demonstrated that MPC outperforms PID control for an artificial network with strong sensory inputs, but the level of control MPC provides might still be lower than what would be needed to fully investigate the structure and function of the neural manifold in a biological circuit. The answer will surely depend on the specific system being studied and the biological questions, but we can suggest two heuristic criteria for evaluating whether a controller is sufficiently effective. First, the controller should produce smaller errors and variability with respect to a target neural trajectory than what is observed across repeated presentations of the original stimulus. In sensory experiments, there is always some variability in the neural or behavioral responses across trials where the same stimulus is presented. Controller-elicited responses should have a variability below the level of the natural sensory responses and regularly result in the desired trajectory. Achieving this would not only demonstrate control efficacy but also offer new insights into the sources of neural and behavioral variability. Second, an effective controller should be capable of identifying a stimulus or stimulus sequence that drives the system toward a desired behavioral or neural state that was not present in the training data used to fit the dynamics model. Success in this regard would indicate meaningful generalization of the latent dynamics and suggest a deeper understanding of the system’s structure.  It is also important to note that the superior performance of MPC comes with increased computational complexity. MPC requires solving an optimization problem at each time step, which can lead to delayed control signals in real-time applications. This added overhead may limit its applicability when ultra-fast control is required or when computational resources are constrained. In contrast, PID control is significantly simpler and faster to implement, making it a more practical choice in scenarios where some loss in control accuracy is acceptable in exchange for speed and efficiency.

Recent years have seen the development of high-density silicon electrodes that can record extracellular spikes from hundreds to thousands of neurons simultaneously \citep{yang2020open,steinmetz2021neuropixels}, but this is still only a tiny fraction of the number of cells that participate in local neural circuits. In Experiment II, we found that the average performance of the forecasting model and the controller decreased as the proportion of neurons observed in the SNN decreased. However, these lower observation conditions also had higher variance in the distribution of average nMSE, indicating that good control over small ensembles is possible. It is also well-known that when observing a subset of states of a dynamical system, time-delay embedding the measurements gives information on the full dynamics \citep{clark_reduced-dimension_2022}. In each of the experiments here, time-delay embedding was not used to fit the VAEs or LDMs for simplicity. Although this did not appear to impact the controller performance for the higher observation percentage models, it may have been detrimental to performance as the percentage decreased. It would be interesting to examine if using time delays would result in increased performance even when the number of observed neurons is very low. Other work has successfully used time-lagged autoencoders to find LDMs \citep{wehmeyer2018time}, and methods exist to find optimal time delays and the number of embedding dimensions \citep{sugihara1990nonlinear}. However, these methods may be too complex and computationally demanding to be used in practice, as recording time is often limited in vivo by electrode drift.

The divide between the descriptive and generative perspectives on neural manifolds is related to the question of whether there is a low-dimensional dynamical system that emerges from the much larger and more complex dynamical system defined by the biophysics of intrinsic and synaptic currents in large populations of neurons. In this study, the true dynamics of the AC were defined by a large, nonlinear dynamical system composed of hundreds of interconnected artificial neurons. We were able to represent the activity of this network in a low-dimensional neural manifold (Figure \ref{fig:LDM_forecasts}) with a clustered structure that is suggestive of attractor basins. The MPC framework developed in this study provided a method for experimentally probing the dynamics on this manifold to better understand its structure and function. Our results are consistent with the generative perspective in that a simple linear approximation of the dynamics in the latent space was sufficient to achieve a high level of control over the network. However, we also found that that there were regions of the state space where the LDM was not a good enough model of the underlying system to provide strong control, a result that can be interpreted as support of the descriptive perspective or as a source of insight for how to improve the latent model. If MPC can be applied in biological systems, it could provide a strong test of whether manifold activity is causal by enabling experimenters to see if specific organismal behaviors can be produced by controlling the latent neural dynamics. This would be an powerful tool in furthering the understanding of how complex behaviors and computations emerge from the structure of neural circuits and the dynamics of their activity. Neural dynamics models as suggested in \citep{sani2021modeling} would be particularly suited for these kinds of tasks since they are able to identify the dynamics directly related to behavior.

It may be important that there was a topological alignment between the latent spaces for neural activity and the stimulus set. The sVAE successfully discovered a dimensional mapping that separated the digits and their variants into 10 distinct and well-separated clusters (Figure \ref{fig:sVAE}A). There was also clear evidence of clustering in the neural latent space that mapped in an orderly way to which digit the stimulus was (Figure \ref{fig:LDM_forecasts}). Though it is beyond the scope of the present study, it would be interesting to explore how the stimulus latent spaces discovered by other dimensionality reduction methods impacts how the LDM performs in forecasting and control. Similarly, alignment has been explored in relating the neural manifolds between organisms on the same tasks. Future work should explore if a controller trained one on organism's manifold can be transferred to another by aligning these latent representations \citep{dabagia2023aligning, ganjali2024unsupervised}. If topological alignment is required for control in this classification task and in other computational problems such as the ones explored by Susillo and Barak \citep{sussillo2013opening}, it may speak to a simple but profound theory that derives from the ideas of James \citep{james1890principles} and Hebb \citep{hebb1949organization}: that learning is a process of aligning the latent dynamics of neural circuits to the latent dynamics of the physical world. MPC, both as theory and as a method for more precise experimental manipulation, may be of benefit in testing this theory in biological systems.
\section{Appendix}

\begin{table}[h!]
    \centering
    \caption{Hyperparameters for neurons in SNN and training on classification task.}
    \begin{tabular}{cccccc}
        \toprule
        $\eta$ & $\beta$ & Trial Length (ms) & Learning Rate & Training Epochs & Batch Size \\
        \midrule
        0.5 & 0.99 & 50 & $5 \times 10^{-4}$ & 33 & 128 \\
        \bottomrule
    \end{tabular}
    \label{tab:SNN_hyperparameters}
\end{table}

\begin{table}[h!]
    \centering
    \caption{Network architecture of the sVAE.}
    \label{tab:svae_architecture}
    \begin{tabular}{lcccccc}
        \toprule
        Layer & Output Size & Kernel Size & Stride & Padding & Activation \\
        \midrule
        \multicolumn{6}{c}{\textbf{Encoder}} \\
        \midrule
        Input & 28 x 28 & -- & -- & -- & -- \\
        Conv Layer 1 & 32 x 28 x 28 & 3 x 3 & 1 & 1 & ReLU \\
        Conv Layer 2 & 64 x 14 x 14 & 3 x 3 & 2 & 1 & ReLU \\
        Conv Layer 3 & 64 x 7 x 7 & 3 x 3 & 2 & 1 & ReLU \\
        Conv Layer 4 & 64 x 7 x 7 & 3 x 3 & 1 & 1 & ReLU \\
        Flatten & 3136 & -- & -- & -- & -- \\
        Linear Layer 1 & 256 & -- & -- & -- & ReLU \\
        Latent Mean ($\mu$) & 2 & -- & -- & -- & -- \\
        Latent Log-Variance ($\log \sigma^2$) & 2 & -- & -- & -- & -- \\
        \midrule
        \multicolumn{6}{c}{\textbf{Decoder}} \\
        \midrule
        Linear Layer 1 & 256 & -- & -- & -- & ReLU \\
        Linear Layer 2 & 3136 & -- & -- & -- & ReLU \\
        Unflatten & 64 x 7 x 7 & -- & -- & -- & -- \\
        ConvTranspose Layer 1 & 64 x 7 x 7 & 3 x 3 & 1 & 1 & ReLU \\
        ConvTranspose Layer 2 & 64 x 13 x 13 & 3 x 3 & 2 & 1 & ReLU \\
        ConvTranspose Layer 3 & 32 x 27 x 27 & 3 x 3 & 2 & 0 & ReLU \\
        ConvTranspose Layer 4 & 29 x 29 & 3 x 3 & 1 & 0 & Sigmoid \\
        \bottomrule
    \end{tabular}
\end{table}
The output of the final deconvolutional layer was trimmed to produce a $28 \times28$ dimensional output image. This architecture was adapted from the example found in \url{https://github.com/rasbt/stat453-deep-learning-ss21/blob/main/L17/1_VAE_mnist_sigmoid_mse.ipynb}.

\begin{table}[h!]
    \centering
    \caption{sVAE training hyperparameters.}
    \begin{tabular}{cccc}
        \toprule
        Learning Rate & Batch Size & Training Epochs & $\alpha$\\
        \midrule
        $5 \times 10^{-4}$ & 128 & 57 & 0.5\\
        \bottomrule
    \end{tabular}
    \label{tab:sVAE_training_hyperparameters}
\end{table}

\begin{table}[h!]
    \centering
    \caption{Network architecture of nVAE. Batch normalization (BN) is applied after each hidden layer before ReLU activation.}
    \label{tab:nVae_architecture}
    \begin{tabular}{lccc}
        \toprule
        Layer & Dimensionality & BatchNorm & Activation \\
        \midrule
        \multicolumn{4}{c}{\textbf{Encoder}} \\
        \midrule
        Input & 500 & -- & -- \\
        Hidden Layer 1 & 100 & \ding{51} & ReLU \\
        Hidden Layer 2 & 50 & \ding{51} & ReLU \\
        Hidden Layer 3 & 10 & \ding{51} & ReLU \\
        Latent Mean ($\mu$) & 2 & -- & -- \\
        Latent Log-Variance ($\log \sigma^2$) & 2 & -- & -- \\
        \midrule
        \multicolumn{4}{c}{\textbf{Decoder}} \\
        \midrule
        Hidden Layer 1 & 10 & \ding{51} & ReLU \\
        Hidden Layer 2 & 50 & \ding{51} & ReLU \\
        Hidden Layer 3 & 100 & \ding{51} & ReLU \\
        Hidden Layer 4 & 500 & \ding{51} & ReLU \\
        Output & 500 & -- & Sigmoid \\
        \bottomrule
    \end{tabular}
\end{table}

\begin{table}[h!]
    \centering
    \caption{nVAE pretraining hyperparameters.}
    \begin{tabular}{cccc}
        \toprule
        Learning Rate & Batch Size & Training Epochs & $\alpha$\\
        \midrule
        $5 \times 10^{-4}$ & 256 & 20 & 0.001\\
        \bottomrule
    \end{tabular}
    \label{tab:nVAE_pretraining_hyperparameters}
\end{table}

\begin{table}[h!]
    \centering
    \caption{LDM training hyperparameters.}
    \begin{tabular}{cccc}
        \toprule
        Learning Rate & Batch Size & Training Epochs & $\alpha$\\
        \midrule
        $5 \times 10^{-4}$ & 256 & 128 & 0.001\\
        \bottomrule
    \end{tabular}

    \label{tab:LDM_training_hyperparameters}
\end{table}

\section*{Data Availability}
All code used for simulation, model fitting, control, and figure generation can be found at: \url{https://github.com/melizalab/Neural-Manifold-MPC.git}.
\subsection*{Acknowledgements}

This work was supported by National Institutes of Health grant 1R01-DC018621, National Science Foundation grant IOS-1942480, and the University of Virginia Brain Institute.


\begin{thebibliography}{10}

\ifx \doiurl  \undefined \def \doiurl#1{\url{https://doi.org/#1}}\fi
\ifx \url     \undefined \def \url#1{\texttt{#1}}\fi
\ifx \burl    \undefined \def \burl#1{\texttt{#1}}\fi
\ifx \etal    \undefined \def \etal{\textit{et al.}}\fi

\bibitem{cunningham2014dimensionality}
John~P Cunningham and Byron~M Yu.
\newblock Dimensionality reduction for large-scale neural recordings.
\newblock {\em Nature Neuroscience}, 17(11):1500--1509, 2014.

\bibitem{mante2013context}
Valerio Mante, David Sussillo, Krishna~V Shenoy, and William~T Newsome.
\newblock Context-dependent computation by recurrent dynamics in prefrontal cortex.
\newblock {\em Nature}, 503(7474):78--84, 2013.

\bibitem{kim2017ringdynamics}
Sung~Soo Kim, Hervé Rouault, Shaul Druckmann, and Vivek Jayaraman.
\newblock Ring attractor dynamics in the drosophila central brain.
\newblock {\em Science}, 356(6340):849--853, 2017.

\bibitem{chaudhuri2019intrinsic}
Rishidev Chaudhuri, Berk Ger{\c{c}}ek, Biraj Pandey, Adrien Peyrache, and Ila Fiete.
\newblock The intrinsic attractor manifold and population dynamics of a canonical cognitive circuit across waking and sleep.
\newblock {\em Nature Neuroscience}, 22(9):1512--1520, 2019.

\bibitem{chung2021neural}
SueYeon Chung and Larry~F Abbott.
\newblock Neural population geometry: An approach for understanding biological and artificial neural networks.
\newblock {\em Current opinion in neurobiology}, 70:137--144, 2021.

\bibitem{langdon2023unifying}
Christopher Langdon, Mikhail Genkin, and Tatiana~A Engel.
\newblock A unifying perspective on neural manifolds and circuits for cognition.
\newblock {\em Nature Reviews Neuroscience}, pages 1--15, 2023.

\bibitem{gallego2017neural}
Juan~A Gallego, Matthew~G Perich, Lee~E Miller, and Sara~A Solla.
\newblock Neural manifolds for the control of movement.
\newblock {\em Neuron}, 94(5):978--984, 2017.

\bibitem{fortunato2023nonlinear}
C{\'a}tia Fortunato, Jorge Bennasar-V{\'a}zquez, Junchol Park, Joanna~C Chang, Lee~E Miller, Joshua~T Dudman, Matthew~G Perich, and Juan~A Gallego.
\newblock Nonlinear manifolds underlie neural population activity during behaviour.
\newblock {\em bioRxiv}, 2023.

\bibitem{pang2016dimensionality}
Rich Pang, Benjamin~J Lansdell, and Adrienne~L Fairhall.
\newblock Dimensionality reduction in neuroscience.
\newblock {\em Current Biology}, 26(14):R656--R660, 2016.

\bibitem{florian2011hidden}
Blaettler Florian, Kollmorgen Sepp, Herbst Joshua, and Hahnloser Richard.
\newblock Hidden markov models in the neurosciences.
\newblock {\em Hidden Markov Models, Theory and Applications}, page 169, 2011.

\bibitem{lusch2018deep}
Bethany Lusch, J~Nathan Kutz, and Steven~L Brunton.
\newblock Deep learning for universal linear embeddings of nonlinear dynamics.
\newblock {\em Nature Communications}, 9(1):4950, 2018.

\bibitem{susillo2015neural_solution_muscle}
David Sussillo, Mark~M Churchland, Matthew~T Kaufman, and Krishna~V Shenoy.
\newblock A neural network that finds a naturalistic solution for the production of muscle activity.
\newblock {\em Nature Neuroscience}, 18(7):1025--1033, 2015.

\bibitem{linderman2017slds}
Scott Linderman, Matthew Johnson, Andrew Miller, Ryan Adams, David Blei, and Liam Paninski.
\newblock Bayesian learning and inference in recurrent switching linear dynamical systems.
\newblock In {\em Artificial intelligence and statistics}, pages 914--922. PMLR, 2017.

\bibitem{rakovic_handbook_2019}
Saša~V. Raković and William~S. Levine, editors.
\newblock {\em Handbook of {Model} {Predictive} {Control}}.
\newblock Control {Engineering}. Springer International Publishing, Cham, 2019.

\bibitem{hewing_learning-based_2020}
Lukas Hewing, Kim~P. Wabersich, Marcel Menner, and Melanie~N. Zeilinger.
\newblock Learning-based model predictive control: Toward safe learning in control.
\newblock {\em Annual Review of Control, Robotics, and Autonomous Systems}, 3(1):269--296, 2020.

\bibitem{brunton_data-driven_2019}
Steven~L. Brunton and Jose~Nathan Kutz.
\newblock {\em Data-driven science and engineering: machine learning, dynamical systems, and control}.
\newblock Cambridge University Press, Cambridge, 2019.

\bibitem{schwenzer_review_2021}
Max Schwenzer, Muzaffer Ay, Thomas Bergs, and Dirk Abel.
\newblock Review on model predictive control: an engineering perspective.
\newblock {\em The International Journal of Advanced Manufacturing Technology}, 117(5-6):1327--1349, 2021.

\bibitem{bieker_deep_2019}
Katharina Bieker, Sebastian Peitz, Steven~L. Brunton, J.~Nathan Kutz, and Michael Dellnitz.
\newblock Deep model predictive control with online learning for complex physical systems.
\newblock {\em arXiv}, 2019.

\bibitem{kaiser_sparse_2018}
E.~Kaiser, J.~N. Kutz, and S.~L. Brunton.
\newblock Sparse identification of nonlinear dynamics for model predictive control in the low-data limit.
\newblock {\em Proceedings of the Royal Society A: Mathematical, Physical and Engineering Sciences}, 474(2219):20180335, 2018.

\bibitem{salzmann_real-time_2023}
Tim Salzmann, Elia Kaufmann, Jon Arrizabalaga, Marco Pavone, Davide Scaramuzza, and Markus Ryll.
\newblock Real-time neural {MPC}: Deep learning model predictive control for quadrotors and agile robotic platforms.
\newblock {\em IEEE Robotics and Automation Letters}, 8(4):2397--2404, 2023.

\bibitem{zheng_physics-informed_2023}
Yingzhe Zheng and Zhe Wu.
\newblock Physics-informed online machine learning and predictive control of nonlinear processes with parameter uncertainty.
\newblock {\em Industrial \& Engineering Chemistry Research}, 62(6):2804--2818, 2023.

\bibitem{frohlich_feedback_2005}
Flavio Fröhlich and Sašo Jezernik.
\newblock Feedback control of {Hodgkin}–{Huxley} nerve cell dynamics.
\newblock {\em Control Engineering Practice}, 13(9):1195--1206, 2005.

\bibitem{yue_non-linear_2022}
Rongting Yue, Ryan Tomastik, and Abhishek Dutta.
\newblock Non-linear model-based control of neural cell dynamics.
\newblock preprint, In Review, 2022.

\bibitem{senthilvelmurugan_active_2023}
Nambi~Narayanan Senthilvelmurugan and Sutha Subbian.
\newblock Active fault tolerant deep brain stimulator for epilepsy using deep neural network.
\newblock {\em Biomedical Engineering / Biomedizinische Technik}, 68(4):373--392, 2023.

\bibitem{fehrman2024nonlinear}
Christof Fehrman and C~Daniel Meliza.
\newblock Nonlinear model predictive control of a conductance-based neuron model via data-driven forecasting.
\newblock {\em Journal of neural engineering}, 21(5):056014, 2024.

\bibitem{bolus_state-space_2021}
M~F Bolus, A~A Willats, C~J Rozell, and G~B Stanley.
\newblock State-space optimal feedback control of optogenetically driven neural activity.
\newblock {\em Journal of Neural Engineering}, 18(3):036006, 2021.

\bibitem{eshraghian2023}
Jason~K. Eshraghian, Max Ward, Emre~O. Neftci, Xinxin Wang, Gregor Lenz, Girish Dwivedi, Mohammed Bennamoun, Doo~Seok Jeong, and Wei~D. Lu.
\newblock Training spiking neural networks using lessons from deep learning.
\newblock {\em Proceedings of the IEEE}, 111(9):1016--1054, 2023.

\bibitem{maass2011liquid}
Wolfgang Maass.
\newblock Liquid state machines: motivation, theory, and applications.
\newblock {\em Computability in context: computation and logic in the real world}, pages 275--296, 2011.

\bibitem{sussillo2013opening}
David Sussillo and Omri Barak.
\newblock Opening the black box: low-dimensional dynamics in high-dimensional recurrent neural networks.
\newblock {\em Neural computation}, 25(3):626--649, 2013.

\bibitem{lecun1998gradient}
Yann LeCun, L{\'e}on Bottou, Yoshua Bengio, and Patrick Haffner.
\newblock Gradient-based learning applied to document recognition.
\newblock {\em Proceedings of the IEEE}, 86(11):2278--2324, 1998.

\bibitem{neftci2019}
Emre~O. Neftci, Hesham Mostafa, and Friedemann Zenke.
\newblock Surrogate gradient learning in spiking neural networks: Bringing the power of gradient-based optimization to spiking neural networks.
\newblock {\em IEEE Signal Processing Magazine}, 36(6):51--63, 2019.

\bibitem{gomari2022variational}
Daniel~P Gomari, Annalise Schweickart, Leandro Cerchietti, Elisabeth Paietta, Hugo Fernandez, Hassen Al-Amin, Karsten Suhre, and Jan Krumsiek.
\newblock Variational autoencoders learn transferrable representations of metabolomics data.
\newblock {\em Communications Biology}, 5(1):645, 2022.

\bibitem{kingma2013auto}
Diederik~P Kingma and Max Welling.
\newblock Auto-encoding variational bayes.
\newblock {\em arXiv}, 2013.

\bibitem{sani2021modeling}
Omid~G Sani, Hamidreza Abbaspourazad, Yan~T Wong, Bijan Pesaran, and Maryam~M Shanechi.
\newblock Modeling behaviorally relevant neural dynamics enabled by preferential subspace identification.
\newblock {\em Nature neuroscience}, 24(1):140--149, 2021.

\bibitem{gosztolai2025marble}
Adam Gosztolai, Robert~L Peach, Alexis Arnaudon, Mauricio Barahona, and Pierre Vandergheynst.
\newblock Marble: interpretable representations of neural population dynamics using geometric deep learning.
\newblock {\em Nature Methods}, pages 1--9, 2025.

\bibitem{schneider2023learnable}
Steffen Schneider, Jin~Hwa Lee, and Mackenzie~Weygandt Mathis.
\newblock Learnable latent embeddings for joint behavioural and neural analysis.
\newblock {\em Nature}, 617(7960):360--368, 2023.

\bibitem{abbaspourazad2024dynamical}
Hamidreza Abbaspourazad, Eray Erturk, Bijan Pesaran, and Maryam~M Shanechi.
\newblock Dynamical flexible inference of nonlinear latent factors and structures in neural population activity.
\newblock {\em Nature Biomedical Engineering}, 8(1):85--108, 2024.

\bibitem{pandarinath2018inferring}
Chethan Pandarinath, Daniel~J O’Shea, Jasmine Collins, Rafal Jozefowicz, Sergey~D Stavisky, Jonathan~C Kao, Eric~M Trautmann, Matthew~T Kaufman, Stephen~I Ryu, Leigh~R Hochberg, et~al.
\newblock Inferring single-trial neural population dynamics using sequential auto-encoders.
\newblock {\em Nature methods}, 15(10):805--815, 2018.

\bibitem{keshtkaran2022large}
Mohammad~Reza Keshtkaran, Andrew~R Sedler, Raeed~H Chowdhury, Raghav Tandon, Diya Basrai, Sarah~L Nguyen, Hansem Sohn, Mehrdad Jazayeri, Lee~E Miller, and Chethan Pandarinath.
\newblock A large-scale neural network training framework for generalized estimation of single-trial population dynamics.
\newblock {\em Nature Methods}, 19(12):1572--1577, 2022.

\bibitem{fiedler_-mpc_2023}
Felix Fiedler, Benjamin Karg, Lukas Lüken, Dean Brandner, Moritz Heinlein, Felix Brabender, and Sergio Lucia.
\newblock do-mpc: {Towards} {FAIR} nonlinear and robust model predictive control.
\newblock {\em Control Engineering Practice}, 140:105676, 2023.

\bibitem{andersson_casadi_2019}
Joel A.~E. Andersson, Joris Gillis, Greg Horn, James~B Rawlings, and Moritz Diehl.
\newblock {CasADi}: a software framework for nonlinear optimization and optimal control.
\newblock {\em Mathematical Programming Computation}, 11(1):1--36, 2019.

\bibitem{wachter_implementation_2006}
Andreas Wächter and Lorenz~T. Biegler.
\newblock On the implementation of an interior-point filter line-search algorithm for large-scale nonlinear programming.
\newblock {\em Mathematical Programming}, 106(1):25--57, 2006.

\bibitem{egner2010expectation}
Tobias Egner, Jim~M Monti, and Christopher Summerfield.
\newblock Expectation and surprise determine neural population responses in the ventral visual stream.
\newblock {\em Journal of Neuroscience}, 30(49):16601--16608, 2010.

\bibitem{Goffinet:2021lowdimensional}
Jack Goffinet, Samuel Brudner, Richard Mooney, and John Pearson.
\newblock Low-dimensional learned feature spaces quantify individual and group differences in vocal repertoires.
\newblock {\em eLife}, 10:e67855, 2021.

\bibitem{khona2022attractor}
Mikail Khona and Ila~R Fiete.
\newblock Attractor and integrator networks in the brain.
\newblock {\em Nature Reviews Neuroscience}, 23(12):744--766, 2022.

\bibitem{adesnik2021probing}
Hillel Adesnik and Lamiae Abdeladim.
\newblock Probing neural codes with two-photon holographic optogenetics.
\newblock {\em Nature neuroscience}, 24(10):1356--1366, 2021.

\bibitem{levina2004maximum}
Elizaveta Levina and Peter Bickel.
\newblock Maximum likelihood estimation of intrinsic dimension.
\newblock {\em Advances in neural information processing systems}, 17, 2004.

\bibitem{chen2022automated}
Boyuan Chen, Kuang Huang, Sunand Raghupathi, Ishaan Chandratreya, Qiang Du, and Hod Lipson.
\newblock Automated discovery of fundamental variables hidden in experimental data.
\newblock {\em Nature Computational Science}, 2(7):433--442, 2022.

\bibitem{song2022modeling}
Christian~Y Song, Han-Lin Hsieh, Bijan Pesaran, and Maryam~M Shanechi.
\newblock Modeling and inference methods for switching regime-dependent dynamical systems with multiscale neural observations.
\newblock {\em Journal of Neural Engineering}, 19(6):066019, 2022.

\bibitem{yang2021adaptive}
Yuxiao Yang, Parima Ahmadipour, and Maryam~M Shanechi.
\newblock Adaptive latent state modeling of brain network dynamics with real-time learning rate optimization.
\newblock {\em Journal of Neural Engineering}, 18(3):036013, 2021.

\bibitem{johnson2016structured}
Matthew~J Johnson, David Duvenaud, Alexander~B Wiltschko, Sandeep~R Datta, and Ryan~P Adams.
\newblock Structured vaes: Composing probabilistic graphical models and variational autoencoders.
\newblock {\em arXiv preprint arXiv:1603.06277}, 2:2016, 2016.

\bibitem{kaiser2021data}
Eurika Kaiser, J~Nathan Kutz, and Steven~L Brunton.
\newblock Data-driven discovery of koopman eigenfunctions for control.
\newblock {\em Machine Learning: Science and Technology}, 2(3):035023, 2021.

\bibitem{yang2020open}
Long Yang, Kwang Lee, Jomar Villagracia, and Sotiris~C Masmanidis.
\newblock Open source silicon microprobes for high throughput neural recording.
\newblock {\em Journal of neural engineering}, 17(1):016036, 2020.

\bibitem{steinmetz2021neuropixels}
Nicholas~A Steinmetz, Cagatay Aydin, Anna Lebedeva, Michael Okun, Marius Pachitariu, Marius Bauza, Maxime Beau, Jai Bhagat, Claudia B{\"o}hm, Martijn Broux, et~al.
\newblock Neuropixels 2.0: A miniaturized high-density probe for stable, long-term brain recordings.
\newblock {\em Science}, 372(6539):eabf4588, 2021.

\bibitem{clark_reduced-dimension_2022}
Randall Clark, Lawson Fuller, Jason~A. Platt, and Henry D.~I. Abarbanel.
\newblock Reduced-dimension, biophysical neuron models constructed from observed data.
\newblock {\em Neural Computation}, 34(7):1545--1587, 2022.

\bibitem{wehmeyer2018time}
Christoph Wehmeyer and Frank No{\'e}.
\newblock Time-lagged autoencoders: Deep learning of slow collective variables for molecular kinetics.
\newblock {\em The Journal ofChemical Physics}, 148(24), 2018.

\bibitem{sugihara1990nonlinear}
George Sugihara and Robert~M May.
\newblock Nonlinear forecasting as a way of distinguishing chaos from measurement error in time series.
\newblock {\em Nature}, 344(6268):734--741, 1990.

\bibitem{dabagia2023aligning}
Max Dabagia, Konrad~P Kording, and Eva~L Dyer.
\newblock Aligning latent representations of neural activity.
\newblock {\em Nature Biomedical Engineering}, 7(4):337--343, 2023.

\bibitem{ganjali2024unsupervised}
Mohammadali Ganjali, Alireza Mehridehnavi, Sajed Rakhshani, and Abed Khorasani.
\newblock Unsupervised neural manifold alignment for stable decoding of movement from cortical signals.
\newblock {\em International Journal of Neural Systems}, 34(01):2450006, 2024.

\bibitem{james1890principles}
William James.
\newblock {\em The Principles of Psychology}, volume~1.
\newblock Dover, New York, 1890.

\bibitem{hebb1949organization}
Donald~O. Hebb.
\newblock {\em The Organization of Behavior}.
\newblock John Wiley and Sons, Incorporated, New York, 1949.

\end{thebibliography}
\end{document}